\documentclass[a4paper,11pt]{article}

\usepackage{color}
\usepackage[dvipsnames]{xcolor}
\usepackage{pos}
\usepackage{caption}
\usepackage{subcaption}
\usepackage{pifont}

\title{Generalization capabilities of neural networks in lattice applications}

\author{S.~Bulusu}
\author*{M.~Favoni}
\author{A.~Ipp}
\author{D.~I.~Müller}
\author{D.~Schuh}

\affiliation[]{Institute for Theoretical Physics, TU Wien, \\
	Wiedner Hauptstr. 8-10, 1040 Vienna, Austria}

\emailAdd{sbulusu@hep.itp.tuwien.ac.at}
\emailAdd{favoni@hep.itp.tuwien.ac.at}
\emailAdd{ipp@hep.itp.tuwien.ac.at}
\emailAdd{dmueller@hep.itp.tuwien.ac.at}
\emailAdd{schuh@hep.itp.tuwien.ac.at}

\abstract{In recent years, the use of machine learning has become increasingly popular in the context of lattice field theories. An essential element of such theories is represented by symmetries, whose inclusion in the neural network properties can lead to high reward in terms of performance and generalizability. A fundamental symmetry that usually characterizes physical systems on a lattice with periodic boundary conditions is equivariance under spacetime translations. Here we investigate the
advantages of adopting translationally equivariant neural networks in favor of non-equivariant ones. The system we consider is a complex scalar field with quartic interaction on a two-dimensional lattice in the flux representation, on which the networks carry out various regression and classification tasks. Promising equivariant and non-equivariant architectures are identified with a systematic search. We demonstrate that in most of these tasks our best equivariant architectures can perform and generalize significantly better than their non-equivariant counterparts, which applies not only to physical parameters beyond those represented in the training set, but also to different lattice sizes.}

\FullConference{%
 The 38th International Symposium on Lattice Field Theory, LATTICE2021
  26th-30th July, 2021
  Zoom/Gather@Massachusetts Institute of Technology
}

\begin{document}
\maketitle

\section{Introduction}

Lattice field theories require an intensive use of numerical simulations, and often supercomputers are employed in order to render such simulations feasible in terms of time consumption. 
The outstanding developments which occurred in the realm of machine learning in the past decade, together with the efforts made to improve hardware technology, have attracted a lot of interest from physicists, who have largely relied on neural networks (NNs) in a wide variety of problems, ranging from condensed matter physics to string theory~\cite{Carrasquilla:2017,Bobev:2020}. Even though these applications have proven to be successful, oftentimes no adaptation of architectures adopted in other fields, e.g. image processing, was put in place. A wise strategy when using tools is to tailor them to meet the requirements of the problem in question. In the case of field theories, a cornerstone is represented by Noether's theorem~\cite{Noether:1918}, which states that for every continuous symmetry of the action there exists a respective conserved current. Therefore, a desirable approach is to design NNs in such a way that the underlying symmetries are respected.
A very important result in this direction has been achieved with the introduction of group equivariant convolutional neural networks (G-CNNs)~\cite{Cohen:2016}, which take care of the preservation of global symmetries, such as translations, rotations and reflections. Also in the context of gauge symmetries there have been recent developments   \cite{Cohen:2019,Kanwar:2020,Boyda:2020,Favoni:2020}.

Lattice field theories are often characterized by invariance under spacetime translations. While in \cite{Cohen:2016} the effectiveness of G-CNNs was shown in computer vision applications, in our recent work~\cite{Bulusu:2021rqz} we focus on the relevance of translational symmetry in lattice field theory. For this discussion, it is sufficient to employ convolutional neural networks (CNNs), which were conceived precisely to include translational equivariance in the network properties. Our goal is to investigate how crucial the preservation of such a symmetry is, with a particular attention to the generalization capabilities of the architectures in terms of different lattice sizes and different physical parameters.

\section{Physical system}
    
In this study we focus on a complex scalar field in 1+1 dimensions with quartic interaction and nonzero chemical potential $\mu$, whose action is the following:
\begin{equation}
    S  = \int \mathrm{d}x_0 \mathrm{d}x_1 \left(  \lvert D_0 \phi \rvert^2 - \lvert \partial_1 \phi \rvert^2  - m^2 \lvert \phi \rvert^2 - \lambda \lvert \phi \rvert^4 \right),
\end{equation}
where $D_0 = \partial_0 - i \mu$, $m$ is the mass and $\lambda$ is the coupling constant. The discretization procedure leads to the action
\begin{equation}
    S_{lat}=\sum_x \left( \eta \lvert \phi_x \rvert^2 + \lambda \lvert \phi_x \rvert^4 -\sum_{\nu = 1}^2 \left( e^{\mu \mspace{2mu} \delta_{\nu, 2}} \phi_x^* \phi_{x + \hat{\nu}} + e^{- \mu \mspace{2mu} \delta_{\nu, 2}} \phi_x^* \phi_{x - \hat{\nu}} \right) \right),
\end{equation}
with $\eta=4+m^2$. The two terms involving the complex conjugation lead to a sign problem that can be eliminated via a dual formulation \cite{Gattringer:2013b}. It maps the field $\phi_x$ into the positive integer field $k_{x,\nu}$ and the integer field $l_{x,\nu}$, where $\nu$ indicates either the temporal or the spatial direction.

\section{Architecture types}

In lattice field theories, the input for the NNs is typically a field configuration, while the output is represented by one or more observables. A translationally invariant architecture produces the same output observables for any shift of the input configuration. Architectures that break the symmetry can learn it only approximately.

\begin{figure}[h]
\centering
\begin{subfigure}[b]{0.45\textwidth}
\includegraphics[width=\textwidth]{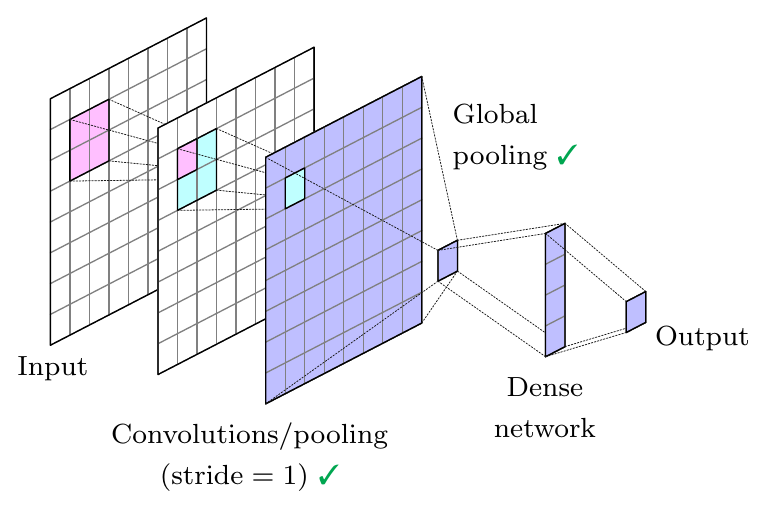}
\caption{Equivariant architecture (EQ)}
\end{subfigure}
\hfill
\begin{subfigure}[b]{0.45\textwidth}
\includegraphics[width=\textwidth]{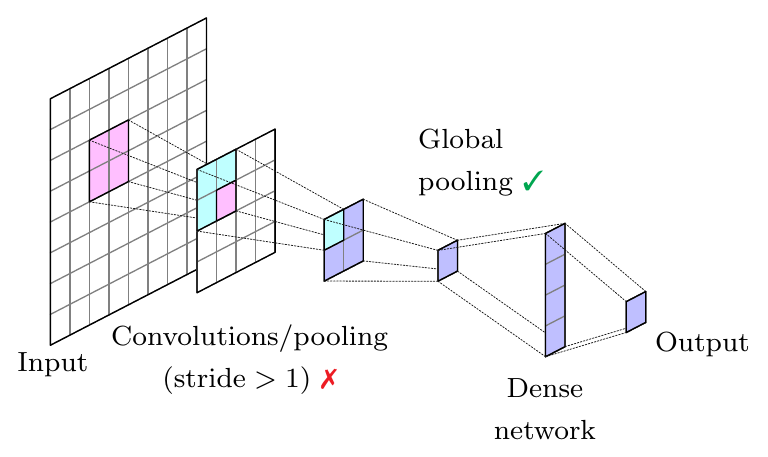}
\caption{Strided architecture (ST)}
\end{subfigure}

\centering
\begin{subfigure}[h]{0.45\textwidth}
\includegraphics[width=\textwidth]{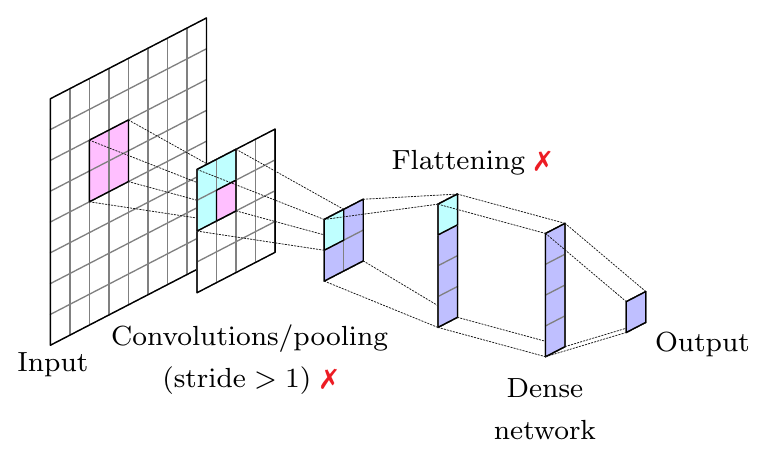}
\caption{Flattening architecture (FL)}
\end{subfigure}
\caption{The architecture types employed to test the relevance of translational symmetry. Checkmarks indicate whether a layer preserves equivariance (\textcolor{Green}{\ding{51}}) or not (\textcolor{Red}{\ding{55}}). A stride of one in the convolutions and in the spatial pooling layers respects translational equivariance (a), while a stride of two or larger breaks it, as in (b) and (c). A flattening layer (c) breaks equivariance and restricts the use of the network to a specific lattice size. Figures from~\cite{Bulusu:2021rqz}.}
\label{fig:archs}
\end{figure}

The general structure we choose for our networks is inspired by typical architectures used in computer vision and can be seen in fig.~\ref{fig:archs}: a first part consisting of several convolutional layers alternated with spatial pooling layers, then a global pooling layer or a flattening step, after which an optional dense network can be appended to make the network more expressive. The type of global pooling depends on the task under examination. For example, the prediction of intensive quantities calls for a global average pooling layer, while global sum pooling is suited for extensive observables. An architecture is guaranteed to be translationally invariant if every layer in the convolutional part of the network is equivariant, meaning that a shift in the input induces a corresponding shift in the output. Another aspect that is worth mentioning is that the theory we study here features periodic boundary conditions, so every layer is equipped with circular padding.

We design three architecture types: the first one (EQ) is translationally equivariant, in the second one (ST) equivariance is broken because of a stride larger than one in the convolutions or in the pooling layers, and in the third one (FL) a flattening step instead of global pooling is used, which also breaks equivariance. ST architectures still retain a residual symmetry based on translations that are a multiple of the stride, but equivariance is lost in general. An additional drawback of the flattening layer is the impediment of employing that architecture on lattice sizes different from the one it has been trained on.

\section{Prediction of observables}

The first task we perform is a regression on two observables, namely the particle number density
\begin{equation}
	n=\frac{1}{N}\sum_xk_{x,2}
\end{equation}
and the average of the modulo squared field
\begin{equation}
    |\phi|^2=\frac{1}{N}\sum_x\frac{W(f_x+2)}{W(f_x)},
\end{equation}
where
\begin{equation}
    f_x=\sum_\nu[|k_{x,\nu}|+|k_{x-\hat{\nu},\nu}|+2(l_{x,\nu}+l_{x-\hat{\nu},\nu})]\,, \quad W(f_x)=\int_0^\infty \mathrm{d}x\, x^{f_x+1}\mathrm{e}^{-\eta x^2-\lambda x^4}.
\end{equation}
These two observables are intensive quantities and therefore we opt for a global average pooling in EQ and ST.

\begin{figure}[h]
    \minipage{0.45\textwidth}
    \centering
    \includegraphics[width=6cm]{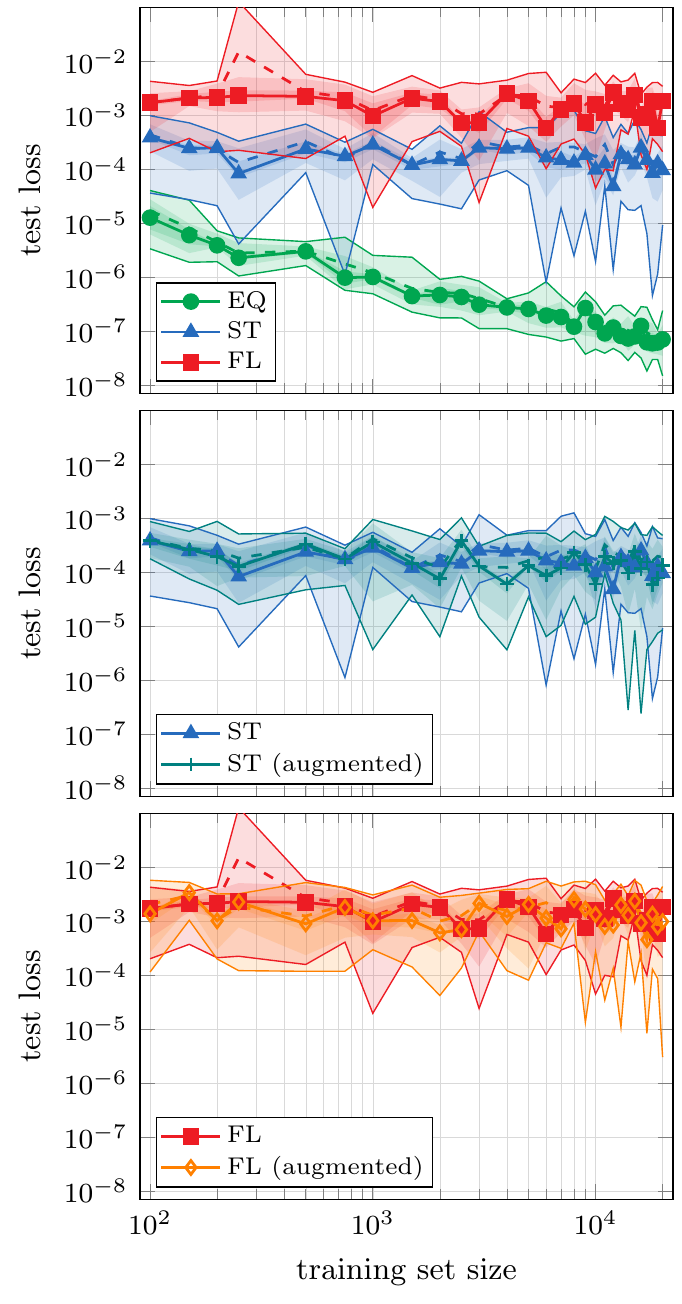}
    \caption{Results of Optuna hyperparameter search. For each architecture type, the model suggested by Optuna is trained for various training set sizes. The bands contain ten instances of such a model, the dashed lines indicate the average test losses and the continuous line passes through the median test losses. The top plot features the comparison of the three architecture types with no data augmentation, the middle and the bottom ones show the test loss behavior with data augmentation for ST and FL, respectively. Image from~\cite{Bulusu:2021rqz}.}
    \label{fig:loss_vs_samples}
    \endminipage
    \hfill
    \minipage{0.45\textwidth}
    \centering
    \includegraphics[width=6cm]{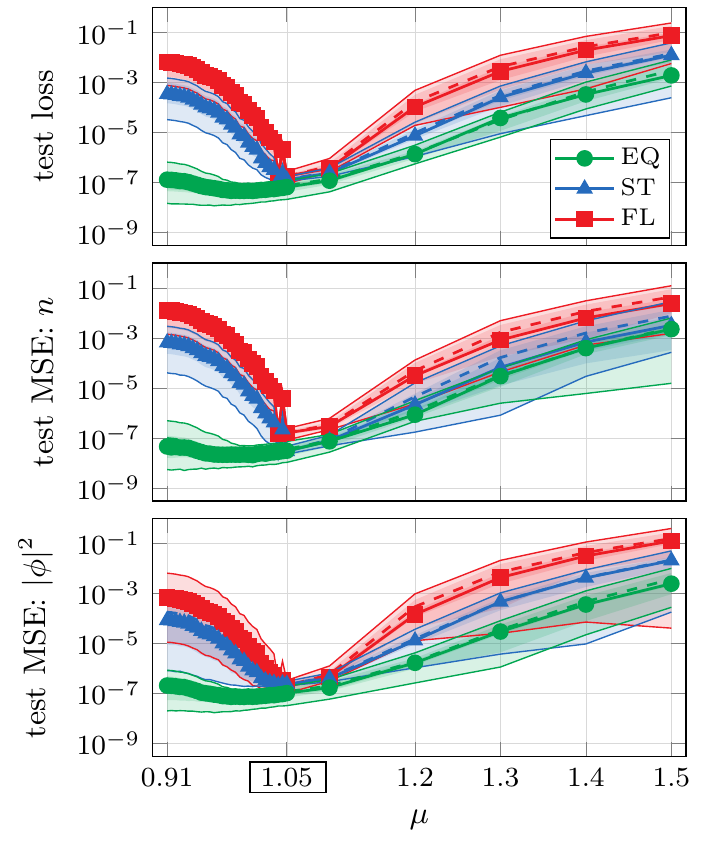}
    \caption{Comparison of the test loss as a function of the chemical potential. The winning architectures in the top plot of fig.~\ref{fig:loss_vs_samples} at the largest training set size are tested on a wide range of chemical potential, while being trained on only one specific value (1.05). Image from~\cite{Bulusu:2021rqz}.}
    \label{fig:loss_vs_mu}
    \endminipage
\end{figure}

The training set is made up of configurations generated with specific values of the physical parameters: $\lambda=1$, $\eta=4.01$, $\mu=1.05$ and lattice sizes $N_t=60$, $N_x=4$.
In the attempt to make the fairest comparison possible, we define for each architecture a large search space for many hyperparameters (e.g.~the number of convolutional layers and of linear layers, the kernel size and the number of channels in each convolution, the position of a spatial pooling layer) and run an automatized optimizer called Optuna~\cite{Akiba:2019} to select the most promising values of such hyperparameters. The criterion to identify the winning hyperparameter combination is the minimum validation loss, which is chosen to be the mean squared error (MSE). Ten instances of the resulting best model of each architecture type are then trained all over again. We repeat the same operation for various training set sizes, ranging from 100 to 20000 samples. After that, we compute the loss on the test set, which consists of 4000 samples for every $\mu\in\{0.91,\dots,1.05\}$ with steps of $\Delta\mu=0.005$. This set is therefore generated with 29 different values of $\mu$, while for training and validating only one was used. This is done because we aim at inspecting the generalization capabilities of the models.

Figure~\ref{fig:loss_vs_samples} shows the outcome of the whole procedure. As is apparent in the top plot, the EQ type performs much better than its non-equivariant counterparts independently of the number of samples used during training. Also, the test loss improves for EQ when increasing the training set size, while for the other two architectures it remains approximately constant. In order to counteract the missing translational equivariance in ST and FL, a new training process is carried out with data augmentation, meaning that configurations are randomly shifted in both directions before being fed into the network. Surprisingly, this does not produce substantial differences in the test loss for ST and FL, as reported in the middle and bottom plot of fig.~\ref{fig:loss_vs_samples}.

\begin{figure}[h!]
    \centering
    \includegraphics{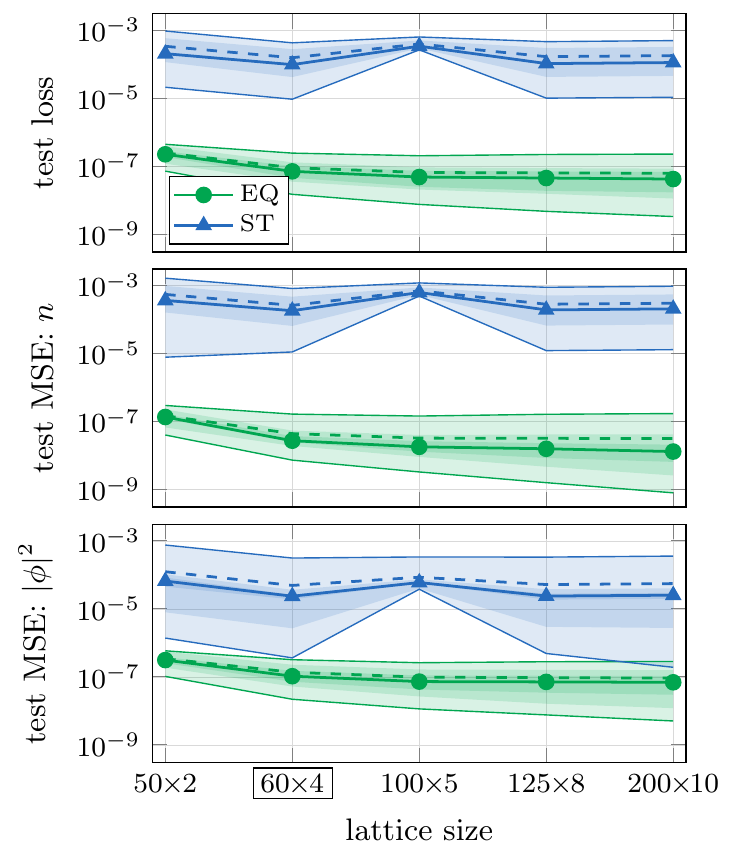}
    \caption{Comparison of the test loss as a function of the lattice size. The best models found at the largest training set size are tested on several lattice sizes, while being trained on only one specific value ($60\times4$). Image from~\cite{Bulusu:2021rqz}.}
    \label{fig:loss_vs_size}
\end{figure}

In the following, we consider only the 10 instances found using 20000 samples during training for each architecture type. In fig.~\ref{fig:loss_vs_mu}, we report the behaviour of the test loss for every individual value of $\mu$. In this analysis we also include  4000 configurations generated at each chemical potential in the range $\mu\in\{1.1,\dots,1.5\}$ with steps of $\Delta\mu=0.1$ to check extrapolation abilities of the models. While ST and FL perform well only where training took place, EQ is able to generalize very well to smaller chemical potentials. The performance deteriorates for all architectures for larger chemical potentials (see a detailed discussion of the reasons in~\cite{Bulusu:2021rqz}), but EQ proves to be a more reliable choice with respect to its non-equivariant counterparts, yielding a test loss smaller by about one order of magnitude.

We now want to examine the generalization to lattice sizes other than the one used for training and validating, so we have to restrict ourselves to a comparison between EQ and ST. Figure~\ref{fig:loss_vs_size} undeniably confirms that for this problem translational equivariance is extremely beneficial, evident from the fact that the loss of EQ is about three orders of magnitude smaller than the loss of ST for all lattice sizes. It is worth mentioning that this task has already been tackled previously in~\cite{Zhou:2019}. There, an architecture of type FL with $\sim10^7$ parameters was trained on the $200\times10$ lattice at two values of the chemical potential and was able to reach a test loss of $10^{-6}$, two orders of magnitude less accurate than our best EQ instance can achieve, even though this last one is trained on a different lattice size on just one value of $\mu$ with much fewer parameters ($\sim10^4)$. Lastly, we point out that the kink in the ST bands is due to another drawback of a stride larger than one: a spatial pooling layer with a stride of two can use the first four rows of the lattice, but discards the fifth one, losing $20\%$ of information, which leads to a much larger loss.

\section{Detection of flux violations}

\begin{figure}
    \centering
    \begin{subfigure}[t]{0.3\textwidth}
    \centering
    \includegraphics[scale=0.8]{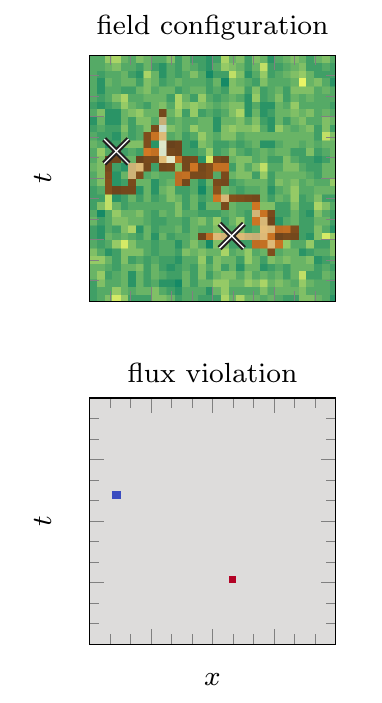}
    \caption{Example field configuration}
    \label{fig:worm}
    \end{subfigure}
    \begin{subfigure}[t]{0.6\textwidth}
    \centering
    \includegraphics[scale=0.8]{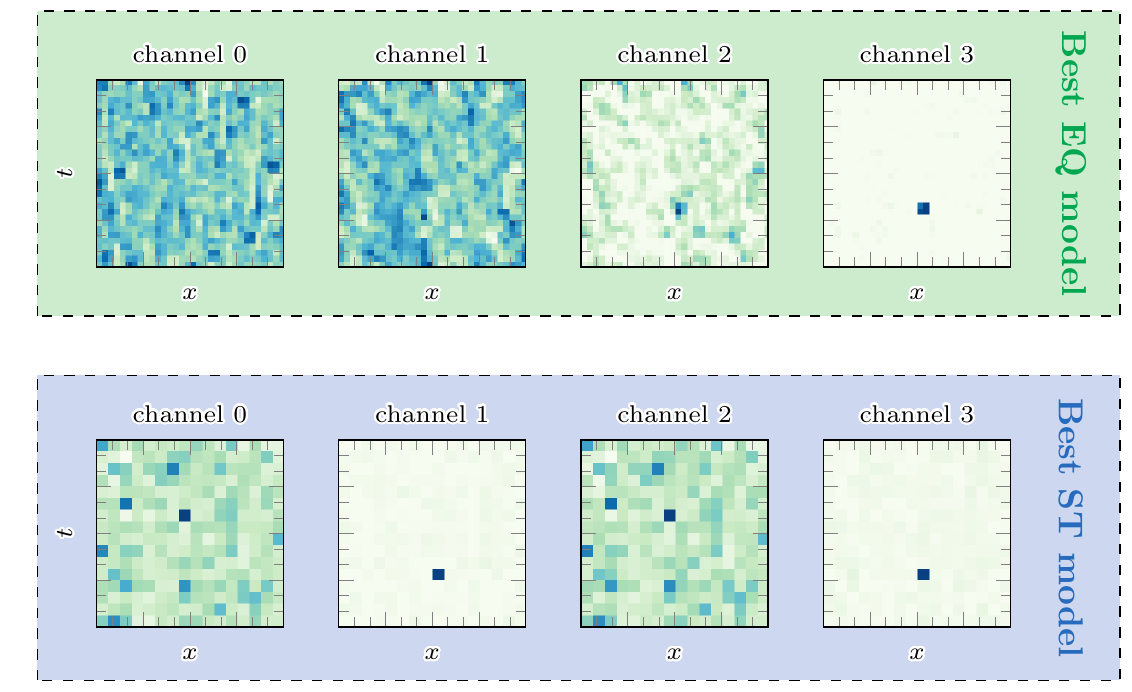}
    \caption{Feature maps of convolutional part in best EQ and ST models}
    \label{fig:flux_detection}
    \end{subfigure}
    \caption{Task visualization and network prediction. The top plot in (a) shows a possible path drawn by the worm algorithm on top of a preexisting physical configuration, while the bottom plot highlights flux violations, which coincide with the worm endpoints. The feature maps in (b) show that successful models detect just one flux violation in only some of the channels in order to discriminate between open and closed worm configurations. Image from~\cite{Bulusu:2021rqz}.}
    \label{fig:flux}
\end{figure}

The dual formulation of the complex scalar field given in sec.~2 is also called flux representation~\cite{Gattringer:2013b}, because the field $k$ obeys the conservation law $\sum_{\nu} \left( k_{x, \nu} - k_{x - \hat{\nu}, \nu} \right) = 0$. Physical configurations must respect this flux conservation, but we can artificially create configurations exhibiting flux violations by slightly modifying the so-called worm algorithm~\cite{Prokofev:2001} used for the data generation in the first task. The algorithm proposes a field update at a particular lattice site; then, if accepted, another update is proposed for one of the next-neighboring sites. The mechanism repeats itself, drawing a path on the lattice that is referred to as a worm, whose head moves until it meets the tail. Before this happens, the ends of the worms violate flux conservation, so we save configurations that feature an open worm. An example of this is depicted in fig.~\ref{fig:flux}.

\begin{figure}[h]
\centering
\begin{subfigure}{0.45\textwidth}
\centering
\includegraphics[width=6.5cm]{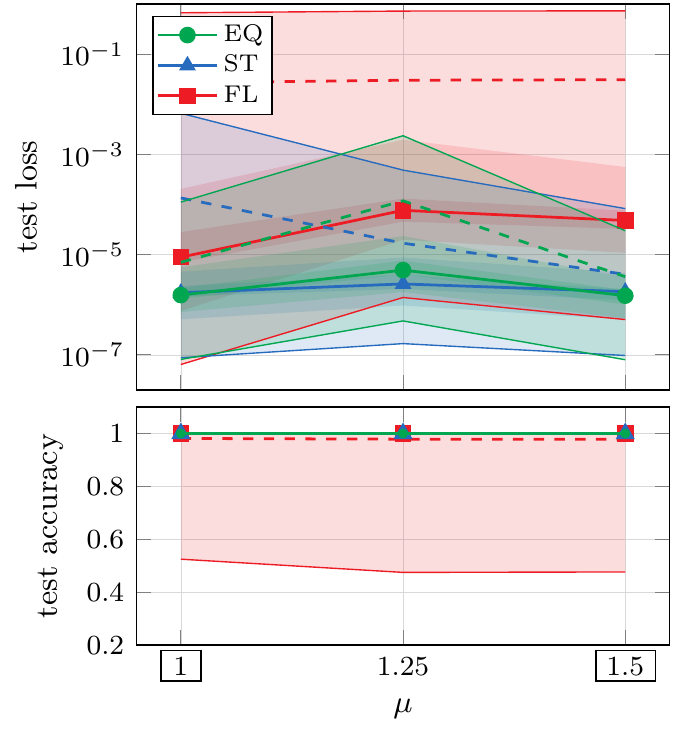}
\caption{Comparison of test loss and test accuracy vs chemical potential on $8\times8$ lattices.}
\label{fig:class_loss_vs_mu}
\end{subfigure}
\hfill
\begin{subfigure}{0.45\textwidth}
\centering
\includegraphics[width=6.5cm]{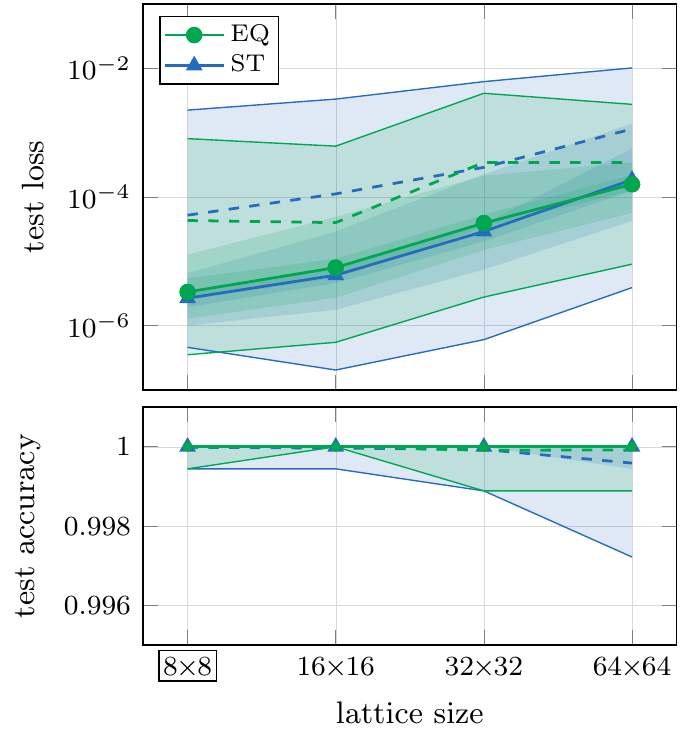}
\caption{Comparison of test loss and test accuracy vs lattice size.}
\label{fig:class_loss_vs_size}
\end{subfigure}
\caption{Results for open worm detection. The bands contain 50 instances of the architectures indicated by Optuna. In (a), the test loss and the test accuracy are reported as functions of the chemical potential on $8\times8$ configurations, while in (b) they are given as functions of the lattice size. Images from~\cite{Bulusu:2021rqz}.}
\label{fig:class_results}
\end{figure}

Apart from $\lambda$, which is kept fixed at 1, the dataset is generated using multiple combinations of physical parameters: $\eta\in\{4.01,4.04,4.25\}$, $\mu\in\{1,1.25,1.5\}$ and lattice sizes $N_t=N_x\in\{8,16,32,64\}$. Only two combinations are used for training, specifically $(\eta,\mu)\in\{(4.01,1.5),(4.25,1)\}$ on $8\times8$ lattices. A total of 4000 samples make up the training set, half of them without open worms, the other half with open worms. The task of the networks is classifying correctly whether a configuration features an open worm or not. Figure~\ref{fig:flux_detection} shows that it is sufficient for the models to detect only one of the flux violations.

Also in this case, an appropriate search space for the hyperparameters is defined and explored using Optuna to suggest the values that minimize the validation loss, which is the binary cross entropy. In a classification task it is not clear what kind of global pooling is the most fitting, which is why this choice is part of the hyperparameter optimization. 50 instances of the best architecture are trained, yielding the results in fig.~\ref{fig:class_results}. The test loss and the test accuracy do not strongly depend on the physical parameters, as we can see in fig.~\ref{fig:class_loss_vs_mu}, where the worse performance of the FL models compared to the other two architecture types is also clearly visible. Figure~\ref{fig:class_loss_vs_size} tells us that the loss deteriorates on larger lattices and that EQ and ST achieve very close results in this classification task.

\section{Counting open worms}

An extension of the previous section can be achieved by adding more than only one open worm on top of a physical configuration. This is then treated as a regression task, which is reminiscent of counting problems such as crowd counting and, with an appropriate adaptation, can be used for the evaluation of $n$-point functions.

\begin{figure}[h]
\begin{subfigure}{0.45\textwidth}
\includegraphics[width=6.5cm]{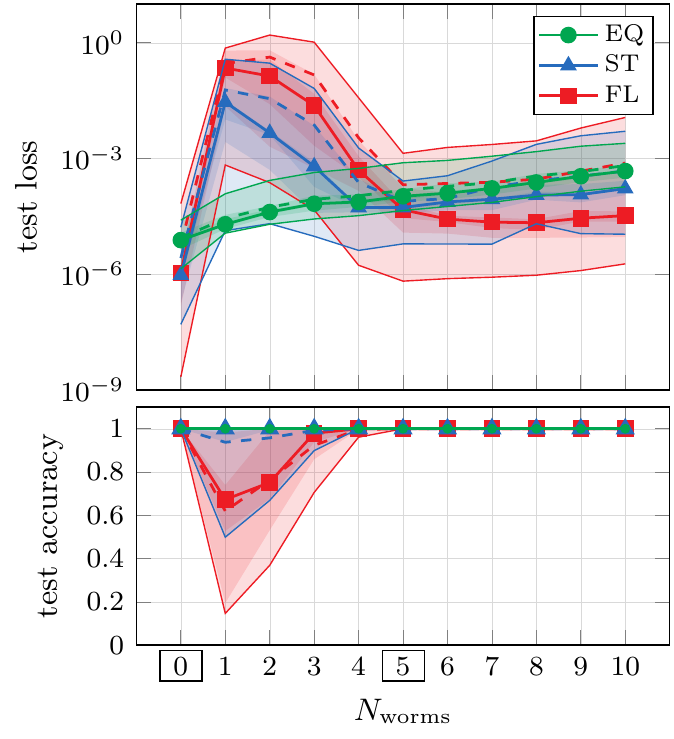}
\caption{Comparison of test loss and test accuracy vs number of open worms on $8\times8$ lattices.}
\label{fig:counting_loss_vs_worms}
\end{subfigure}
\hfill
\begin{subfigure}{0.45\textwidth}
\includegraphics[width=6.5cm]{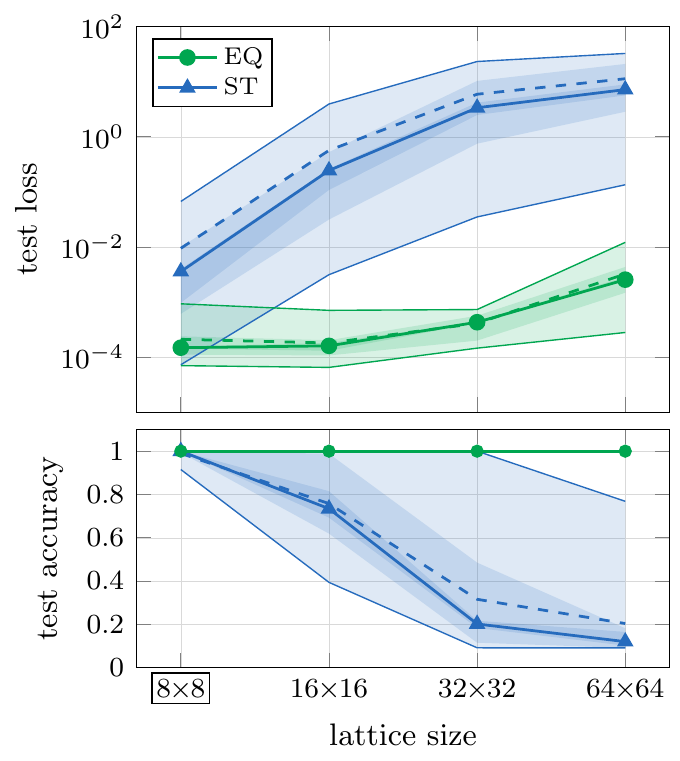}
\caption{Comparison of test loss and test accuracy vs lattice size.}
\label{fig:counting_loss_vs_size}
\end{subfigure}
\caption{Results for worm counting. The bands contain 20 instances of the architectures indicated by Optuna. In (a), the test loss and the test accuracy are reported as a function of the open worms on $8\times8$ configurations, whereas in (b) they are given as functions of the lattice size. Images from~\cite{Bulusu:2021rqz}.}
\label{fig:counting_results}
\end{figure}

The combinations of physical parameters are inherited from the previous section with the addition of a number of open worms ranging from 0 to 10. For training we will use the two combinations employed in the classification task with 0 and 5 worms. This amounts to only four combinations out of the 396 possible ones. The number of training samples is 20000, and also in this case we optimize the hyperparameters by means of Optuna. Since we are dealing with an extensive quantity, we adopt a global sum pooling.

Figure~\ref{fig:counting_loss_vs_worms} reports the test loss and accuracy on $8\times8$ lattices as a function of open worms, where ST and FL perform worse than EQ. In particular, they have a hard time predicting a number of worms from 1 to 4. In fig.~\ref{fig:counting_loss_vs_size} the results on every lattice are shown, indicating that EQ has to be preferred, even achieving $100\%$ test accuracy with all of its 20 instances.

\section{Conclusions}

We have tested the performance of three architecture types on three different tasks. The architectures that respect translational symmetry proved to be a highly reliable choice in all tasks, while the architectures that break translational invariance can feature a poor performance when it comes to generalizing to physical parameters that were absent in the training set. Moreover, breaking invariance brings other drawbacks along: if a stride larger than one is used, part of the input information can be lost, while a flattening layer hinders the possibility of applying the same architecture to other lattice sizes. Another important aspect to keep in mind is that the automated optimizer we have used, Optuna, has favored small or medium-sized architectures, with a number of trainable parameters between $10^2$ and $10^5$.
In conclusion, our study shows that it is sensible to use translationally equivariant neural networks for problems that are characterized by translational symmetry.

\acknowledgments

This work has been supported by the Austrian Science Fund FWF No.~P32446-N27, No.~P28352 and Doctoral program No.~W1252-N27. The Titan\,V GPU used for this research was donated by the NVIDIA Corporation.

\bibliographystyle{JHEP.bst}
\bibliography{references.bib}

\end{document}